\documentclass[aps,pre,twocolumn,superscriptaddress]{revtex4-1}
\usepackage[dvips]{graphicx}
\usepackage{amssymb,amsmath,bm}
\usepackage{here}
\usepackage{color}
\newcommand{\ei}{\epsilon_{\rm{in}}}
\newcommand{\eo}{\epsilon_{\rm{out}}}
\newcommand{\bei}{\overline{\epsilon_{\rm{in}}}}
\newcommand{\beo}{\overline{\epsilon_{\rm{out}}}}

\begin{document}


\title{Dynamics of assembly production flow}


\author{Takahiro Ezaki}
\email{ezaki@jamology.rcast.u-tokyo.ac.jp}
\affiliation{Department of Aeronautics and Astronautics, Graduate School of Engineering, The University of Tokyo, 7-3-1 Hongo, Bunkyo-ku, Tokyo 113-8656, Japan}
\affiliation{Japan Society for the Promotion of Science, \\8 Ichibancho, Kojimachi, Chiyoda-ku, Tokyo 102-8472, Japan}

\author{Daichi Yanagisawa}
\altaffiliation[Present address: ]{Department of Aeronautics and Astronautics, Graduate School of Engineering, The University of Tokyo, 7-3-1 Hongo, Bunkyo-ku, Tokyo 113-8656, Japan}
\affiliation{Department of Aeronautics and Astronautics, Graduate School of Engineering, The University of Tokyo, 7-3-1 Hongo, Bunkyo-ku, Tokyo 113-8656, Japan}

\author{Katsuhiro Nishinari}
\affiliation{Research Center for Advanced Science and Technology, The University of Tokyo, 4-6-1 Komaba, Meguro-ku, Tokyo 153-8904, Japan}


\date{\today}

\begin{abstract}
Despite recent developments in management theory, maintaining a manufacturing schedule remains difficult because of production delays and fluctuations in demand and supply of materials. 
The response of manufacturing systems to such disruptions to dynamic behavior has been rarely studied. To capture these responses, 
we investigate a process that models the assembly of parts into end products. 
The complete assembly process is represented by a directed tree, where the smallest parts are injected at leaves and the end products are removed at the root. 
A discrete assembly process, represented by a node on the network, integrates parts, which are then sent to the next downstream node as a single part.
The model exhibits some intriguing phenomena, including overstock cascade,  phase transition in terms of demand and supply fluctuations, nonmonotonic distribution of stockout in the network, and the formation of a stockout path and stockout chains. 
Surprisingly, these rich phenomena result from only the nature of distributed assembly processes. From a physical perspective, 
these phenomena provide insight into delay dynamics and inventory distributions in large-scale manufacturing systems.
\end{abstract}


\maketitle

\section{introduction}
The twentieth century saw industrialized societies develop with the support of organized modern manufacturing,  accompanied by a rapid increase in demand and consumption of goods. Scientific management of production systems dates back to the early twentieth century \cite{Taylor}. 
After Taylor's pioneering work, innumerable studies have been undertaken to control, optimize, and predict production flow in factories \cite{Croom,Dallery1992}. These studies have contributed to the design and management of manufacturing systems. 
However, the description of production flows by these theories, including the queuing theory, remains unsatisfactory, especially for complex, dynamic production systems \cite{Papa1996,Govil1999}. A pivotal factor impeding our understanding of production flows is the dynamic properties of a complex production system as a ``many-body system.'' In general, constraints pertaining to the volume of components and  the finite capacity (buffer) for each job make the system dynamics complex.

This paper proposes a physical approach---suitable for dealing with dynamic phenomena---to the complex system dynamics. 
We use a simple model that captures the essence of the dynamics and makes it {\it visible}.
In the context of nonequilibrium statistical physics, the asymmetric simple exclusion process (ASEP) \cite{Ligget1985,Spohn1992,Derrida1998} has been vigorously studied as the most archetypal model of particle flow with the exclusion (blocking) effect, and nontrivial behaviors specific to  nonequilibrium systems have been discussed. 
In addition, the connection between the queuing theory and ASEP has been reported in recent studies \cite{Arita2009, Yana2010}. 
Thus, we find the intersection of studies on manufacturing systems and physics.
Note that in the supply-chain management field, it is known that even a simple system can lead to highly complex behavior, including chaos \cite{Mosek1988,Hwarng2008} and the so-called {\it bullwhip effect} \cite{Lee1997,Mett1997,Chen2000,Ouy2007,Ouy2010}, which has also been investigated from a physical perspective \cite{Helbing2003,Nagatani2004,Helbing2004,Helbing2004-2}.

Hopp and Spearman \cite{Hopp2000} attempted the idea of using a physical approach called {\it factory physics} to study production flow.
They successfully systemized the fundamental knowledge about production systems with simple mathematics and static analyses. 
However,
recent physical approaches using stability analyses \cite{Helbing2003,Nagatani2004,Helbing2004,Helbing2004-2}  and 
the formulation presented in this paper use a dynamic treatment, and are thus beyond the scope of the factory physics concept. 
The main goal of the present study is to uncover and understand dynamic phenomena observed in complex manufacturing systems that are beyond intuitive prediction. For this purpose, simple models are used to identify the correlation between phenomena and causes. 
Different from 
recent similar attempts \cite{Helbing2003,Nagatani2004,Helbing2004,Helbing2004-2}, 
 we use a full discrete model---space, time, and inventories are discretized---to realize this approach.
This enables us to understand the system at the microscopic level, disregarding relatively less important factors
 (e.g., input and output buffers at each job,  and prediction and adaptation mechanisms),
and to focus on the nature of ``assembly.''

In this paper, we focus on a set of jobs in a single manufacturing unit that constitutes a supply chain. 
Each job corresponding to materials' processing does not predict its supply and demand, but follows the state of neighboring jobs.
In the absence of prediction, the bullwhip effect does not occur. 
The primary goal of this study is to reveal how ``assembly'' processes on a large production line affect the overall system. The system is perturbed by three
types of fluctuations: demand fluctuations, supply fluctuations, and removal  of defective products.
The assembly process involves merging two or more production flows, where material provisions from these streams synchronize (couple) \cite{Baccelli1989}. This significantly increases complexity when the system is under perturbations, thus making rigorous analysis of such system very challenging. 

In previous studies, a set of assembly processes has been represented abstractly by a network of jobs, buffers (nodes), and topology of parts flow (links) \cite{Nof1997,Adan1989,Papa1996,Baccelli1989,Dallery1997}. 
Here, we reformulate an assembly system as an interacting particle system \cite{Ligget1985} and observe its dynamic aspects.
The collective behavior of exclusive particles moving in a discrete network has received considerable attention from physicists \cite{Embley2009,Basu2010,Neri2011,Ezaki2012,Mottishaw2013
}.
However, the effect of particle coalescence (assembly) on these systems has not yet been fully understood.

The reminder of the paper is organized as follows.
The the focal model definition is given in the next section.
To understand the model in detail, we first focus on the dynamics of a restricted parameter set (Sec. \ref{fun}) and then consider variations in the fixed parameters (Sec. \ref{var}). Finally, we summarize the results and discuss the outlook of the study in Sec. \ref{dis}. 

\section{Model}\label{mod}
\begin{figure*}[tbp]
   \includegraphics[width=170mm]{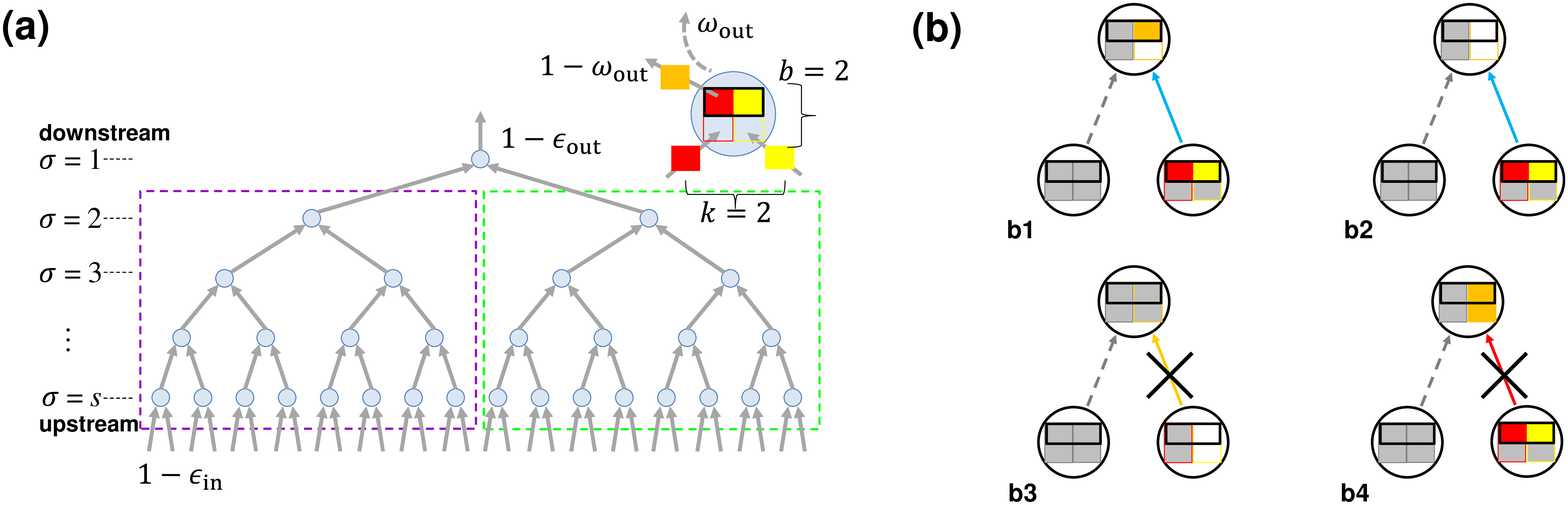}
 \caption{(a) Example of the assembly network ($k=2,s=5$). 
We define blocks-A and -B as the blocks surrounded by purple (left) and green (right) lines, respectively.
\textcolor{black}{(b) Local transition rules for $b=2$ and $k=2$. When the focal node (at right bottom) has both types of parts and the buffer in its downstream node is not full (b1 and b2), it can send a part (production state). If even one required part is absent (b3), the node does not send a part (stockout state). When the node is not in the stockout state (that is, it is capable of producing a part) and the buffer in the next node is full (b4) it does not send a part (demand-deficiency state).}}
 \label{system}
\end{figure*}

Consider a directed regular tree network whose indegree and outdegree are $k$ and $1$, respectively (Fig. \ref{system}). Parts (particles) are transported along links, and at each node, $k$ different parts are assembled, generating a part for the next node. 
A node has a buffer of size $b$ for each incoming part; that is, a node can contain $b$ parts of one type at the same time. 
Hence, the stock status at a node is described by a set of parts for the $k$ buffers. 
The network has $s\in \mathbb{N}$ assembly stages, resulting in the need for $k^{s}$ raw materials. 
To investigate the statistical properties of large-scale production systems, we assume that $k^{s}$ is large.

Basic rules of particle transportation in the system are as follows:
at each time step, an assembly node sends a part to the next node if and only if all the required parts corresponding to the incoming links are stored in its buffers, and the buffer in the next node for the product the current node is creating is not full. Here, this node state is referred to as the ``production state.'' If even one part required in the production is absent, the node does not send a part. This node state is referred to as the ``stockout state.'' Even if the necessary materials are available, the node cannot send the product to the next node if the buffer of the next node is full. This state is referred to as the ``demand-deficiency state.'' In this paper, these three states comprise the ``operational state.''
Noted that, regardless of the state, each node can accept a part only if the corresponding buffer is not full.
These procedures are performed in parallel at each node. 
Thus, the transition of a particular type of parts in a node in stage-$\sigma$ ($\sigma =2,3,\cdots,s-1$) is summarized as follows. In the production state, the 
transition is $-1$ if the part is not provided by its upstream process; otherwise, it is $\pm 0$. In the stockout or demand-deficiency state, the transition is $\pm 0$ if the part is not provided by its upstream process; otherwise, it is $+ 1$. 
At stage-$s$ (the network ``leaves''), $k$ parts are provided, each with probability $1-\epsilon_{\rm{in}}$. The end product is removed at stage-$1$ (the ``root'') with probability $1-\epsilon_{\rm{out}}$. The probabilities $\epsilon_{\rm{in}}\in [0,1)$ and  $\epsilon_{\rm{out}}\in [0,1)$ are the error rate of supply and demand, respectively. These rates are related to the intensity of their fluctuations.

The model is later generalized by introducing a defect rate $\omega_{\rm{out}}\in [0,1)$ reflecting the 
 possibility of producing a defective product, which is immediately abandoned and not sent to the next node.
When a node is in the production state, the produced part is sent to the next node with probability $1-\omega_{\rm{out}}$; otherwise, it is removed from the system as a defective part.

\section{Fundamental phenomena}\label{fun}
In this section, we restrict ourselves to considering the case of 
$(\omega_{\rm{out}}=0,b=2,k=2)$ as the most fundamental example.
These restrictions will be relaxed later.
In the absence of the possibility of producing defective parts, parts flow in each link is conserved.

The case of $b=2$ is important for the following reasons. First, this is the minimum size that allows a stationary production flow. If each node can hold only one inventory item, products can be sent once per two steps as in ASEP. Second, having surplus stock in each process is costly. Generally, in the production management field, reducing stock has been encouraged \cite{Krafcik,Womack,Shah}.
Given cost considerations and conventional domain recommendations, the buffer size we have set is ideal if supply fluctuations are negligible.  Third, setting the buffer size to $2$ leads to a unique stationary state of production; that is, when demand and supply are steady, all nodes are in the production state with a single inventory item in each buffer. This state is robust for small fluctuations in supply, which will be discussed later.

\subsection{Overstock cascade}
\begin{figure*}[tbhp]
   \includegraphics[width=120mm]{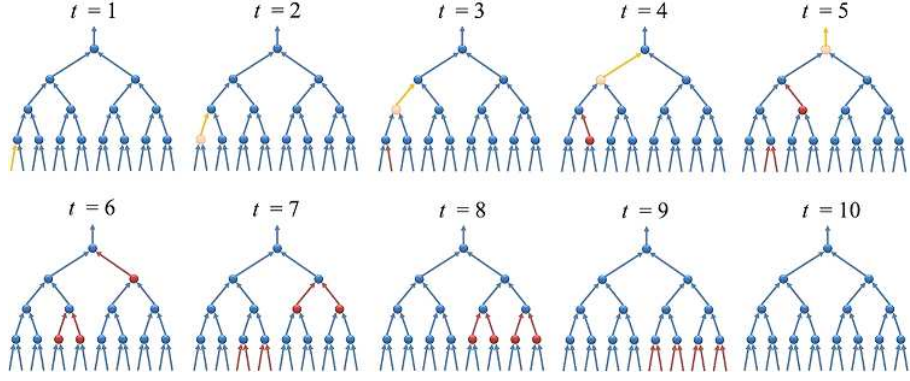}
 \caption{Overstock cascade. Blue, yellow, and red circles denote nodes in production, stockout, and demand-deficiency states, respectively.
 Blue arrows indicate parts flow at the time step; other arrows indicate absence of flow caused by stockout or demand-deficiency.
 We consider a situation where the supply and demand of parts at the boundaries are stationary ($\epsilon_{\rm{in}}=\epsilon_{\rm{out}}=0$).
 Here, we show the response of the system to a supply failure at the leftmost node in stage-$4$.
 The stockout propagates downstream, generating demand-deficiency for the different branches. In contrast, the demand-deficiency state propagates upstream.
 After 10 time steps, the uniform stationary state is recovered.
 }
 \label{fig:prop}
\end{figure*}

First, we focus on the $\epsilon_{\rm{out}}=0$ and small $\epsilon_{\rm{in}}$ cases
to understand the fundamental system behavior.
Figure \ref{fig:prop} illustrates the propagation of the stockout state and the generation of overstock (full  buffers).
For the stationary state, where all nodes are in the production state and their buffers have one part each,
the delay of a single part is input. 
A stockout caused by  supply fluctuation travels downstream (from $\sigma=s$ to $1$). 
A node in the stockout state induced by a shortage of the focal part stops its production, 
while the rest of the parts are provided to the node as scheduled, resulting in the buffers becoming full.
For these excess inventories, upstream processes stop production. This is observed as 
the emergence of the demand-deficiency state, which propagates upstream. This is in contrast with the stockout state because all inventories
are increased by one at the demand-deficiency node in the next step. This leads to a demand-deficiency state at its upstream nodes.
These local dynamics result in an ``overstock cascade.''  
After these disruptions, each node recovers to the stationary production state in the presence of steady input and output. 
In addition, pair annihilation occurs when the stockout state and the demand-deficiency state encounter each other.
At this point, the demand-deficiency state spreads downward, with the exception of the link corresponding to the stockout. These states cancel each other out because  overstocking is avoided due to the shortage of incoming parts.
Here, the stockout state acts as a time buffer for overstocking.
In addition, when two stockout states converge at the same node, they behave as a single state after the merger. Thus, these interactions decrease the number of stockout-state nodes.
\subsection{Production rate}\label{steep}
Here, we investigate the production rate $q$, defined as the expected output value of end products in one time step.
Note that because each part is used once for one end product, the flow of parts in each link is identical to the production rate. 

\begin{figure*}[tbp]
   \includegraphics[width=140mm]{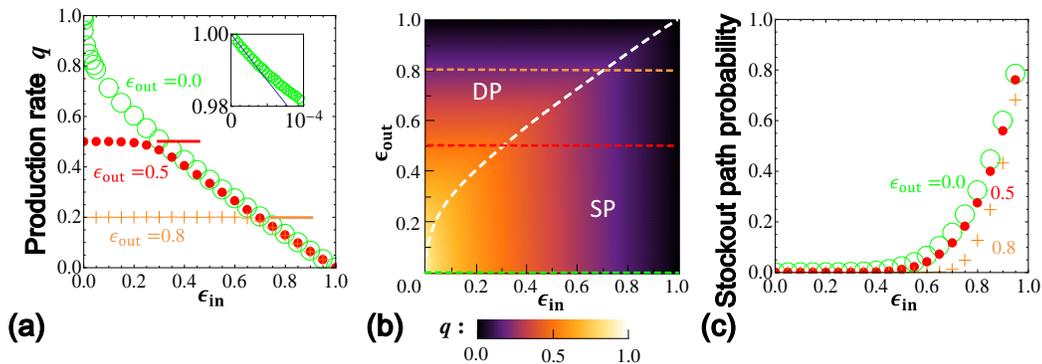}
 \caption{(a) Production rate for various error rates. 
The expression for plateaus observed in DP, $q=1-\epsilon_{\rm{out}}$, is illustrated with bold lines. The blue line in the inset corresponds to Eq. (\ref{lin}).
(b) Phase diagram. When $\epsilon_{\rm{in}}$ ($\epsilon_{\rm{out}}$) is relatively small, the production flow is restricted by the demand (supply), leading to DP (SP). The phase boundary is \textcolor{black}{schematically} indicated with a white dashed line. (c) Stockout-path probability\textcolor{black}{, which is defined as the probability of the system having at least one stockout path.}}
 \label{fig:q}
\end{figure*}

Figure \ref{fig:q}(a) shows the production rate for various error rates. Here, we set $s=8$.
The most remarkable characteristic is the steep decay of $q$ for $\epsilon_{\rm{out}}=0$ and small $\epsilon_{\rm{in}}$, which is asymptotic to the linear expression 
\begin{equation}
q\simeq 1 - k^{s}\epsilon_{\rm{in}}. \label{lin}
\end{equation}
When the error rate $\epsilon_{\rm{in}}$ is sufficiently small, the effects of two different stockouts and their propagations 
do not interact, i.e., the occurrence probability of this interaction is negligible compared
with $\epsilon_{\rm{in}}$.
Hence, a part shortage generated at leaves with rate $k^{s}\epsilon_{\rm{in}}$ directly reduces the production rate.
This expression is illustrated in the inset in Fig. \ref{fig:q}(a).
Since $k^{s}$ is a large number, the production rate decreases significantly, even with a small error supply rate. After this steep drop, $q$ slopes
gently downward because the interaction between the stockout and demand-deficiency nodes is no longer negligible, 
 pair annihilation and coalescence occur, and the supply errors do not directly decrease the flow.
These characteristics stem from the effect of finite buffer capacity (i.e., particle exclusion). When there is no limitation on buffer size, the production rate is expressed as $q = 1-\epsilon_{\rm{in}}.$ (See Appendix \ref{infbuff} for its derivation.)

When $\epsilon_{\rm{out}}$ is nonzero, the production rate $q$ has a plateau of $q = 1-\epsilon_{\rm{out}}$, and the production flow is limited by demand. In other words, in this region, nodes are almost fully occupied with inventories, and wait for the output that takes place with a probability of $1-\epsilon_{\rm{out}}$. After this plateau, the flow becomes limited by supply errors near a certain $\epsilon_{\rm{in}}$. As such, the system has two different phases, the demand-limited phase (DP) and the supply-limited phase (SP), which are shown in Fig. \ref{fig:q}(b). 
In each phase, the production rate depends on only the corresponding error rate\textcolor{black}{; and thus 
the curves in Fig. \ref{fig:q}(a) merge into the case of $\epsilon_{\rm{out}}=0.$}
In DP, the bottleneck limiting the flow is clearly demand (output point of the end product). However, the problem is not as simple in SP. There are many supply inputs and production paths, and the bottleneck that limits the flow contains a few stockout nodes. Moreover, these limiting nodes move temporally. 
\textcolor{black}{Thus an analytical description of the flow is very challenging.}

\subsection{Distribution of the stockout state}
\begin{figure}[tbp]
   \includegraphics[width=80mm]{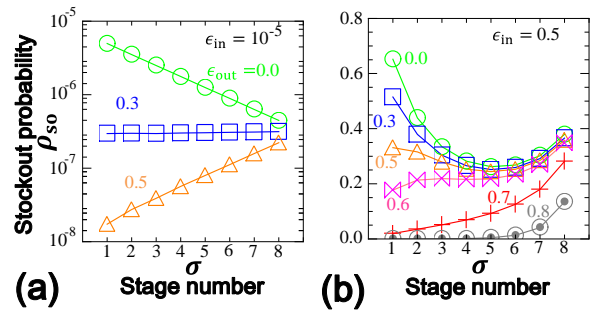}
 \caption{Stockout probability in each stage for (a) small supply-error rate $\epsilon_{\rm{in}} = 10^{-5}$ (exponential behavior Eq. (\ref{eq:r})) and (b) large 
 supply-error rate $\epsilon_{\rm{in}} = 0.5.$ (nonmonotonic relationship).}
 \label{fig:dis}
\end{figure}

For each stage, we study the probability of finding a node in the stockout state, $\rho_{\rm{so},\sigma}$. 
Here, we do not focus on the demand-deficiency state found with probability $\rho_{\rm{dd},\sigma}$ because a complementary relationship exists: $\rho_{\rm{dd},\sigma} = 1 - q - \rho_{\rm{so},\sigma}$. 
Moreover, the production state can be found with probability $q$, irrespective of the stage number.
These probabilities characterize the situations in each stage.
Simulation results are shown in Fig. \ref{fig:dis}.

For a small supply error $\epsilon_{\rm{in}}=10^{-5}$, where interaction among delay propagations is negligible, 
the distribution is monotonic and follows
\begin{equation}
\rho_{\rm{so},\sigma}\propto [k(1-\epsilon_{\rm{out}})^2]^\sigma.\label{eq:r}
\end{equation}
This can be understood by the following approximate evaluation, which ignores the probabilities of $O(\epsilon_{\rm{in}})$.
 In DP with small $\epsilon_{\rm{in}}$ ($\ll \epsilon_{\rm{out}}$), each node is
 in either state-$\textrm{\boldmath $1$}$ or state-$\textrm{\boldmath $2$}$; all buffers have one part (state-$\textrm{\boldmath $1$}$) or are fully occupied (state-$\textrm{\boldmath $2$}$) because the occurrence of a stockout that breaks uniformity is nearly negligible.
Note that these states are defined only by buffers of a focal node and are independent of that node's operational state, which takes the next node's state into account.
After producing an assembled part, the node assumes state-$\textrm{\boldmath $1$}$; i.e., state-$\textrm{\boldmath $1$}$ appears with probability 
$q = 1 - \epsilon_{\rm{out}}$. Hence state-$\textrm{\boldmath $2$}$ can be found with probability $1-q = \eo$.
We then consider the propagation of the stockout state generated at leaves. 
The probability of finding the incoming stockout state to a node in stage-$\sigma$ is expressed as $k\rho_{\rm{so},\sigma+1}q (1-\eo)$ by ignoring the correlation between the nodes. 
The factor $k\rho_{\rm{so},\sigma+1}$ represents the probability of having an incoming stockout state, and $q$ is the probability of finding state-$\textrm{\boldmath $1$}$ at the focal node. If its next node is in state-$\textrm{\boldmath $2$}$ (which state is observed with the probability $\eo$), the propagating stockout state vanishes. Thus, the factor $q (1-\eo)$ reflects the probability that the stockout state can survive in the node.
On the other hand, the probability of a stockout going out of the node is simply $\rho_{\rm{so},\sigma}$.
Hence, by equating them, we obtain $\rho_{\rm{so},\sigma} = k(1-\epsilon_{\rm{out}})^2\rho_{\rm{so},\sigma+1}$.
For example, for $k=2$, $\epsilon_{\rm{out}} =1- \frac{1}{\sqrt{2}}\simeq 0.2929$ gives uniform distribution of the stockout state. (See Fig. \ref{fig:dis}(a) $\eo=0.3.$)
In addition, when $\epsilon_{\rm{out}}=0$, the nodes are nearly in state-$\textrm{\boldmath $1$}$, and
by ignoring $O(\epsilon_{\rm{in}}^2)$, the probability of a node entering the stockout state
is given by $k\rho_{\rm{so},\sigma+1}$, which is included in Eq. (\ref{eq:r}).
These evaluations agree with the simulation results. Note that these arguments are not applicable to  general values of $b$.

The distribution shows an intriguing characteristic when the supply-error rate is large ($\epsilon_{\rm{in}} = 0.5$).
The stockout probability $\rho_{\rm{so}}$ has a local minimum around stage-$5$ for cases in SP, while it is monotonic in DP. 
In SP, successive nodes are observed in the stockout state (hereafter ``stockout chain'').

In the nodes providing parts to the stockout chain, the demand deficiency is induced at each time step. Because the demand-deficiency state is generated upstream, a stockout state that appears in the network leaves vanishes as it proceeds downstream. This is why $\rho_{\rm{so},\sigma}$ increases from $\sigma = 6$ to $8$. In contrast, once the stockout state reaches downstream stages,
it becomes difficult to eliminate and tends to stay as a stockout chain, thereby increasing its probability of remaining because the number of demand deficiencies generated in downstream nodes decreases.
In this phase, the demand deficiency generated by the output error rate $\epsilon_{\rm{out}}$ at stage-$1$ is less dominant than that generated by the supply error $\epsilon_{\rm{in}}$.
 For this reason, the stockout probability $\rho_{\rm{so},\sigma}$ increases as the stage approaches completion, i.e., $\rho_{\rm{so},\sigma}$ drops as $\sigma$ increases. The relationship between $\epsilon_{\rm{out}}$ and the amount of increase near the root (stage-$1$) can be understood by
considering the following equation
\begin{equation}
q = (1-\rho_{\rm{so},1})(1-\eo).\label{qandeo}
\end{equation} 
Since in SP, $q$ is independent of $\eo$, the stockout probability must increase to $\rho_{\rm{so},1} =1 - q  (1-\eo)^{-1}$ at this point. 
Note that Eq. (\ref{qandeo}) is satisfied for general $\ei, \eo, k, b$, and $s$. 
Even if $\omega_{\rm{out}}$ is nonzero, the modified equation, \textcolor{black}{
$q = (1-\rho_{\rm{so},1})(1-\eo)(1-\omega_{\rm{out}})$ holds.}
What is important in this figure is that \textcolor{black}{for a fixed $\epsilon_{\rm{in}}$,} the states in upstream processes are only slightly affected by $\epsilon_{\rm{out}}$ and downstream processes in SP. Hence, a high stockout probability for downstream nodes is not essential to ensure a high production rate; however, it is effective in reducing the stockout at leaves, for example, by extending their buffers. This finding provides a new perspective for the buffer allocation problem \cite{Buzacott,Hillier,Powell,ShiL,Smith2005,ShiC}.

\subsection{Stockout path and inventory imbalance}
With a large supply error, the production flow is dominated by the stockout state path (stockout chain) running from the first to the final stage (hereafter ``stockout path''). When a certain part is not provided to a leaf node for a number of time steps, its downstream processes run out of these parts, while buffers for other parts in the system become full.
Once the stockout path is formed, it tends to remain at the same location for the following reasons.
While waiting for the focal part, inventories are accumulated in the other nodes, and as a result, when the required part is provided, the focal path consumes it immediately, once again exhausting the part. 
The probability of finding this stockout path in the system (stockout path probability; SPP) is shown in Fig. \ref{fig:q}(c), where the stockout path becomes salient when $\epsilon_{\rm{in}}>0.5$ for relatively small $\epsilon_{\rm{out}}$, and  $\epsilon_{\rm{in}}>0.6$ for large $\epsilon_{\rm{out}}$. Judging from the similarity between $\epsilon_{\rm{out}}=0.0$ and $0.5$ in this region of $\epsilon_{\rm{in}}$ (SP), it is supposed that SPP is nearly independent of demand fluctuations. However, while the system is in DP and for sufficiently large $\epsilon_{\rm{out}}$, the increase in SPP is impeded by excessive inventories caused by small demand.

\begin{figure*}[tbp]
   \includegraphics[width=160mm]{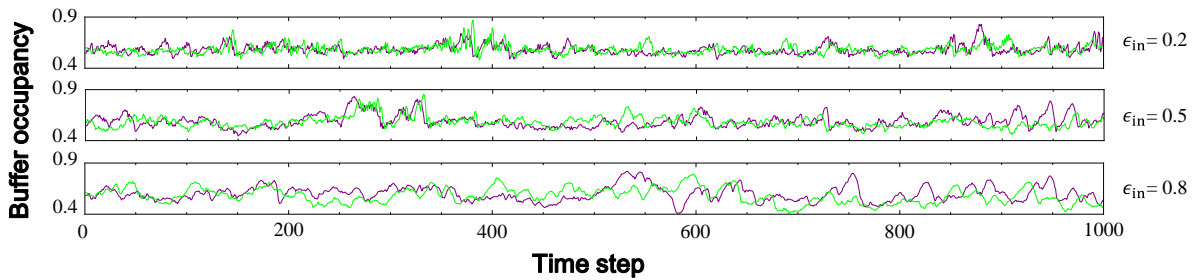}
 \caption{Time series of buffer occupancy in block-A (purple) and block-B (green) for $\epsilon_{\rm{out}}=0$ and different values of $\epsilon_{\rm{in}}$ ($0.2, 0.5, 0.8$). As $\epsilon_{\rm{in}}$ increases, the two lines decouple.}
 \label{fig:ts}
\end{figure*}

We plotted the time series of buffer occupancy of the left and right halves of the system (Fig. \ref{system}) in Fig. \ref{fig:ts}. For small supply fluctuations, two time-series transitions behave similarly; however, as the fluctuation increases, they desynchronize. Although the production rate is undermined through this decoupling, note that the average stock level remains largely unchanged.

\section{Effects of network size, buffer size, and product inspection}\label{var}
Now, we focus on the dependence of variations in  $k, b, s,$ and $\omega_{\rm{out}}$ on the system dynamics.

\subsection{Degree $k$ and buffer size $b$}
\begin{figure*}[tbp]
   \includegraphics[width=160mm]{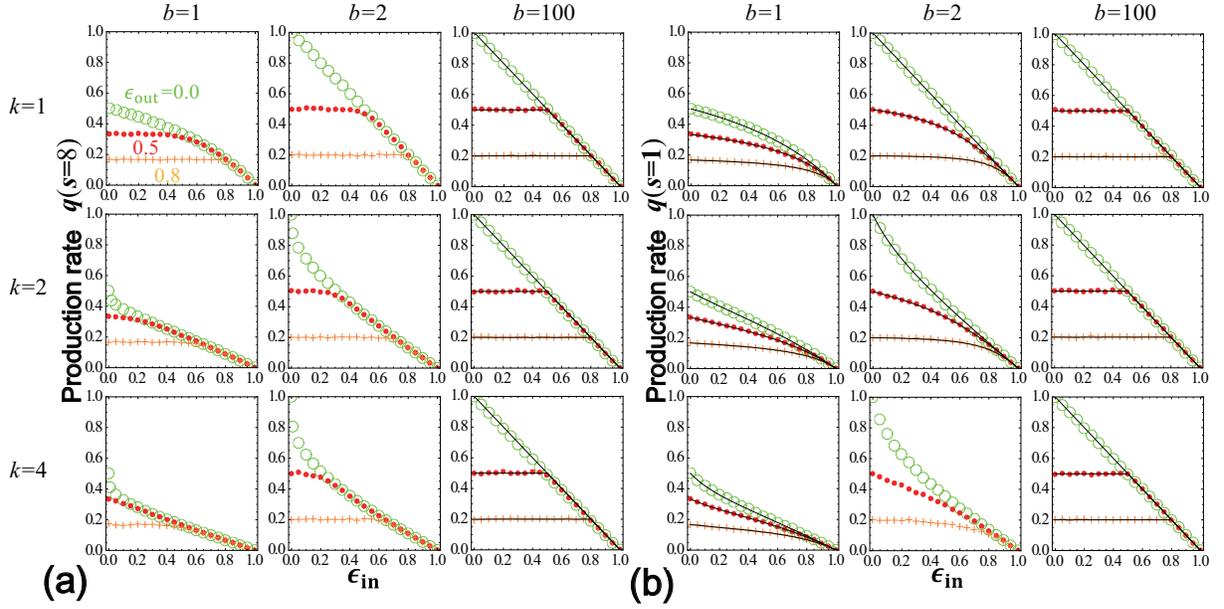}
 \caption{Flux diagrams for various degrees $k$ and buffer sizes $b$. The defect rate is set to $\omega_{\rm{out}}=0.$
For $b=100$, theoretical predictions are given by Eq. (\ref{bufinf}), while for $b=1$ and $b=2$, analytical expressions are derived in Appendices \ref{asmall1} and \ref{asmall2}, respectively.
}
\label{ssone}
\end{figure*}

Keeping the other parameters fixed as $s=8$ and $\omega_{\rm{out}} =0$, we discuss here the 
variations in degree $k$ and buffer size $b$. Figure \ref{ssone}(a) shows the flux diagrams for $k=1,2,$ and $4$ and $b=1,2,$ and $100$ cases (and thus the center panel in each block is the same as in Fig. \ref{fig:q}).
Note that each $k=1$ case (shown for reference) corresponds to a chain of production processes and does not have assembly processes.  
When $b=1,$ the system coincides with ASEP. 

Generally, a too small buffer size ($b=1$) for mitigating exclusion (blocking) effects leads to the exacerbation of particle flux. 
A large difference in fluxes can be observed between $b=1$ and $2$ because $b=2$ is the minimum  buffer that allows stationary flow.
The mitigation of  the exclusion effect explains the sharpness of the phase transitions between SP and DP for larger $b$.
A large buffer is also effective for mitigating the effect of disruption in particle flow, and thus contributes to increasing the flow. 

Degree $k$ is directly connected to the steepness of flux at $\epsilon_{\rm{in}}=0$ [Fig. \ref{ssone}(a)] as discussed in Sec. \ref{steep}.
When the buffer size is large enough to absorb disruptions, the flux diagrams become independent of $k$: 
\begin{equation}
q = \min \left\{ {1-\epsilon_{\rm{in}}, 1-\epsilon_{\rm{out}}}\right\}\label{bufinf}
\end{equation}
See Appendix \ref{infbuff} for its derivation.

Figure \ref{pso_rho1} displays the stockout probability $\rho_{\rm{so},\sigma}$, and the particle density is defined as
\begin{equation}
\rho_{\sigma}\equiv \frac{\langle \tau^t_\sigma\rangle}{b},
\end{equation}
where $\tau^t_{\sigma}$ denotes the number of particles in a buffer in
a node in stage-$\sigma$. In the presence of system symmetry, this value is independent of the buffer position.
As $s$ increases, the particle density increases, absorbing 
fluctuation in inventory arrivals and thereby reducing the stockout probability. However, compared with the large number of inventories 
a node has to retain, the effect is not significant. 
This implies that a policy of increasing buffers is relatively ineffective for increasing the production rate. Rather, a focus on synchronization of the inventory level between nodes is more relevant to achieving cost-effective operations.

\begin{figure*}[tbp]
   \includegraphics[width=160mm]{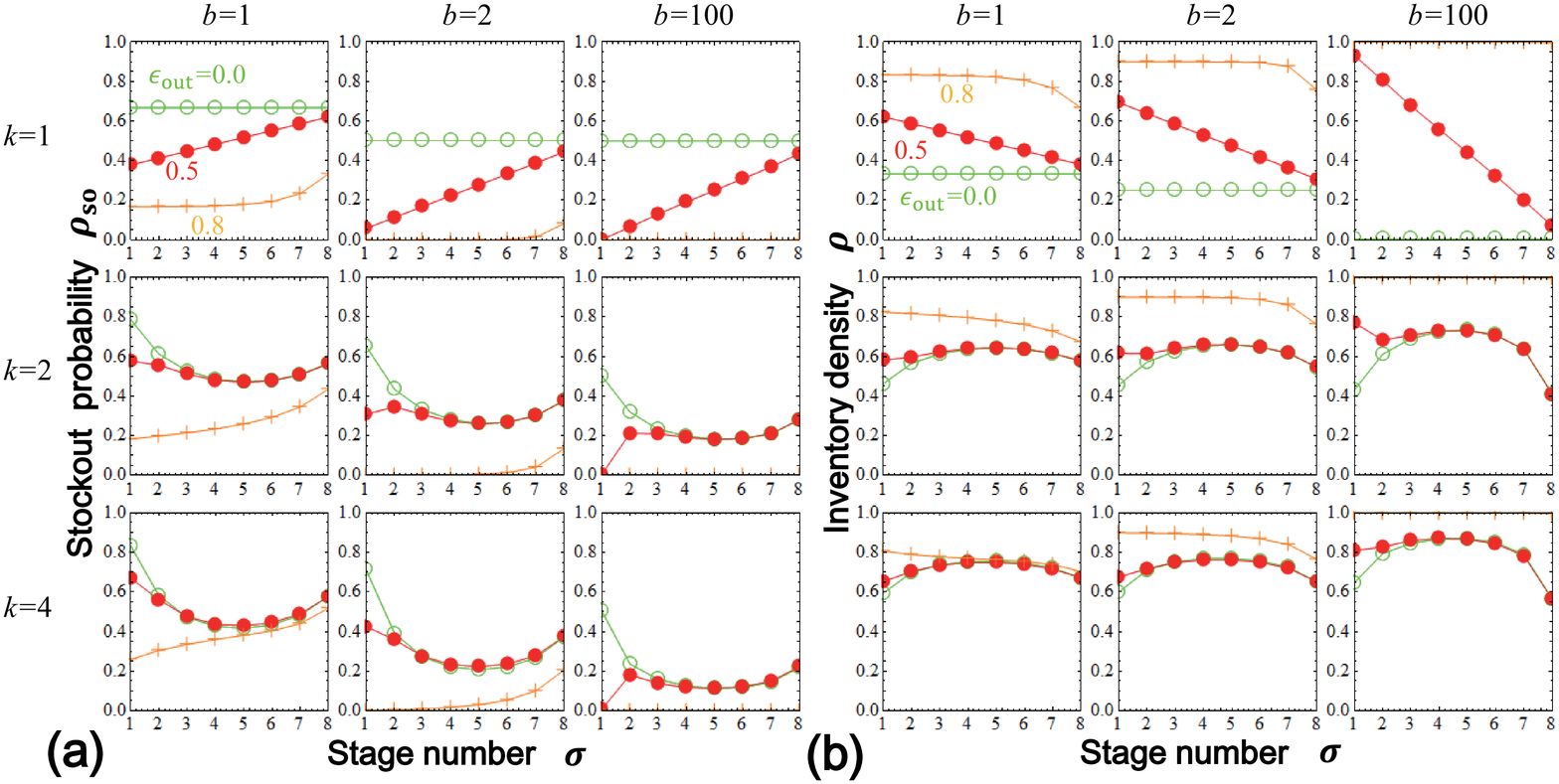}
 \caption{(a) Stockout probability and (b) density distribution for various degrees $k$ and buffer sizes $b$. The other parameters are fixed as $\ei=0.5, s=8$, and $\omega_{\rm{out}}=0.$}
\label{pso_rho1}
\end{figure*}

\subsection{Stage number $s$}
By investigating $s=1$ cases, where assembly is conducted only once in the production process, and comparing the results with $s=8$ cases, we discuss the effect of the stage number $s$.
Figure \ref{ssone}(b) shows  flux diagrams for $s=1$. 
Note that, because in these cases, the system is small enough, the flux can be calculated analytically (see Appendices \ref{asmall1} and \ref{asmall2}). 
Stage number reduction directly leads to the total number of provision errors at supply boundaries,
increasing the particle flux when the supply side is limiting the flow.  
On the other hand, when $\eo$ is more dominant, flux is deteriorated compared with $s>1$ cases, because 
buffers in each stage contribute to absorbing fluctuations that originate in supply errors, and thus
plateaus of $q=\eo$ disappear when the buffer size and $\eo$ are small.

\subsection{Defect rate $\omega_{\rm{out}}$}
\begin{figure*}[tbp]
   \includegraphics[width=160mm]{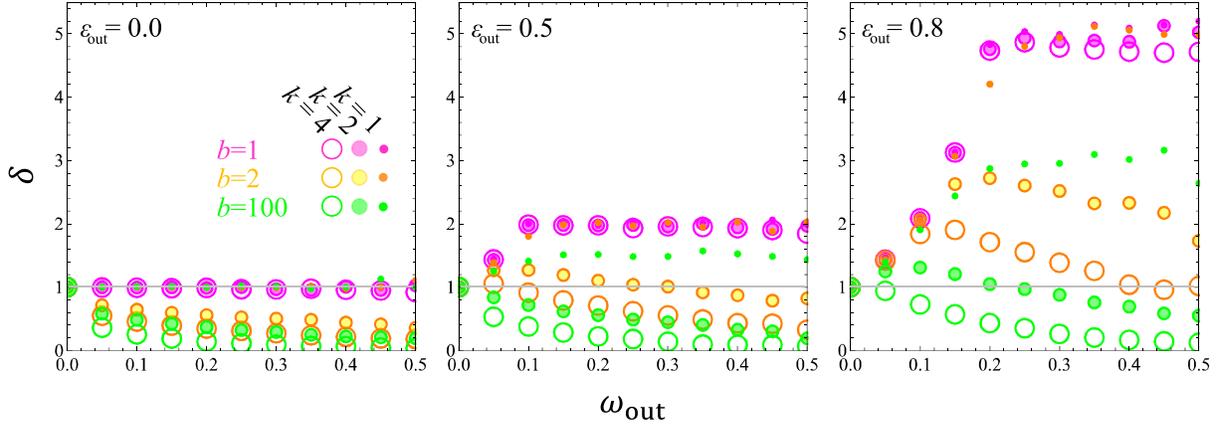}
 \caption{Difference from the expected flux deterioration due to particle removal, defined by Eq. (\ref{delta}). The parameters are set as $\epsilon_{\rm{in}}=0.5$ and $s=8$. 
}
\label{fig:buwatto}
\end{figure*}

In the presence of positive $\omega_{\rm{out}}$, particle flow is jeopardized $s$ times by defects generated in the production process. Hence, when particles flow smoothly, the flux is reduced to $(1-\omega_{\rm{out}})^s$ times. Figure \ref{fig:buwatto} shows relative differences from this ratio, defined as
\begin{equation}
\delta = \frac{q(\omega_{\rm{out}})}{q(0)}[(1-\omega_{\rm{out}})^s]^{-1}\label{delta},
\end{equation}
where $q(0)$ denotes the production rate when the defect rate is zero. 
These figures indicate that
increasing $k$ ($b$) strengthens (weakens) the effect of defects. 
This is because large $k$ increases the randomness of parts provision to downstream nodes, which decreases 
effective production rates, while large $b$ contributes to mitigating this randomness, especially when 
demand is restricted (i.e., a sufficient number of parts are stocked in downstream nodes).
Interestingly, these effects are highly dependent on $\omega_{\rm{out}}$. When $\omega_{\rm{out}}$ is small,
particles are not often removed, and the system is demand-driven. In this state, inventories are well stocked in downstream nodes, reinforcing the effect of buffers. 
In contrast, when $\omega_{\rm{out}}$ is large, many particles are removed before reaching the end 
of the tree, and the buffer effect is no longer present. This explains the non-monotonic behavior of $\delta$ in Fig. \ref{fig:buwatto}.

\begin{figure*}[tbp]
   \includegraphics[width=160mm]{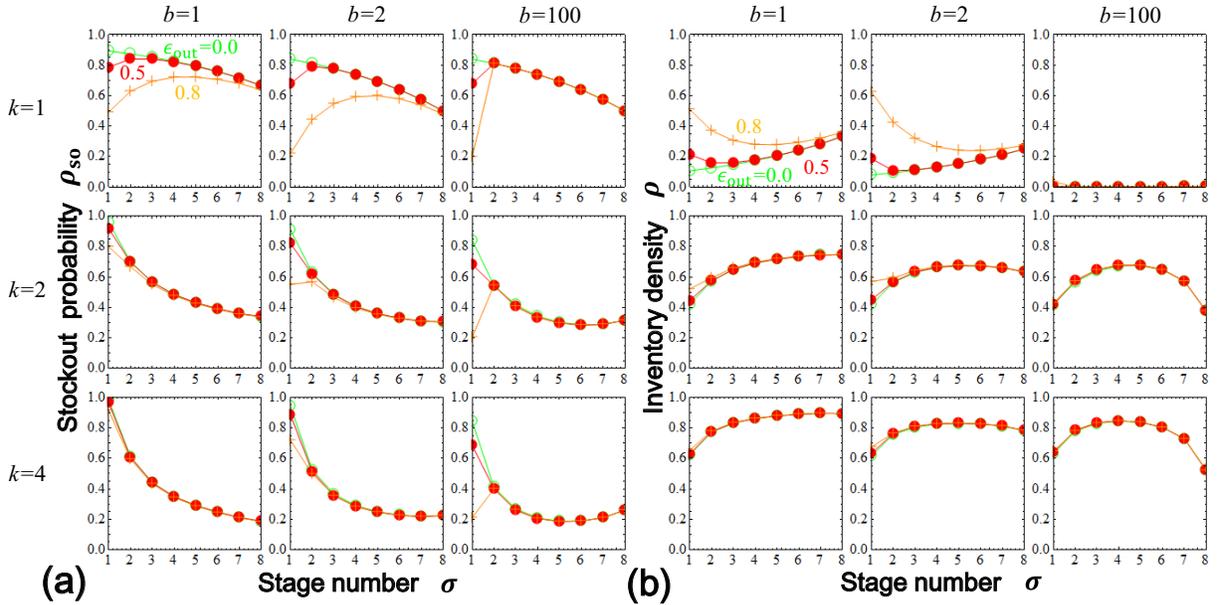}
 \caption{(a) Stockout probability and (b) density distribution for various degrees $k$ and buffer sizes $b$. The other parameters are fixed as $s=8$ and $\omega_{\rm{out}} = 0.15.$}
\label{omegad}
\end{figure*}

Figure \ref{omegad} illustrates the stockout probability and the density distribution for $\omega_{\rm{out}}=0.15$.
This is a sufficiently large value for the particle removal process to determine the system dynamics. 
Hence, when $k\geq 2$, these distributions are almost independent of $\eo$ and determined mainly by network structure and buffer size. Note that for a large buffer size, they coincide with the $\eo=0$ cases in Fig. \ref{pso_rho1}, which is consistent with the fact that fluctuations caused by particle removal can be absorbed 
by large buffers. The effect of particle removal is more significant in downstream stages (small $\sigma$) since particles become scarce as they approach completion. 
For smaller values of $\omega_{\rm{out}}$, 
the aforementioned characteristics do not change drastically between Figs. \ref{pso_rho1} and \ref{omegad}.

\section{Discussion}\label{dis}
We modelled the transportation of parts in a factory, and revealed fundamental characteristics of production flow that are caused by the effects of exclusion and coalescence of particles under supply, demand, and particle removal fluctuations.

First, we confirmed the propagation of a stockout and demand deficiency caused by a single supply error \textcolor{black}{(Fig. \ref{fig:prop})}. With this knowledge, we then analyzed the characteristics of the end product production rate and found steep decay for small supply-error rates \textcolor{black}{(Fig. \ref{fig:q}(a))}. In general, the system behavior can be categorized into two distinctive phases depending on its output and input error rates: DP and SP \textcolor{black}{(Fig. \ref{fig:q}(b))}. 

In SP, production flow is restricted by a moving bottleneck, i.e., successive stockout nodes (the stockout chain).  For larger error rates, the production rate is mostly controlled by the stockout path, i.e., the stockout chain penetrates all stages, which tends to stay at the same location. 
\textcolor{black}{Corresponding to this stockout path, the inventory level in blocks of the production network largely fluctuates (Fig. \ref{fig:ts}).
This inventory imbalance can be eased by the control policy. For example, stopping production if a node is producing more parts than the current bottleneck process could mitigate symmetry breaking. This would reduce cost for holding inventories and can stabilize the system. Also note that this prediction is consistent with the spirit of the {\it kanban} production system \cite{Sugimori,Ohno,Junior}, in which upstream processes are permitted to produce only when they receive orders from downstream processes.
}

In addition, for input errors that are not small---the order of ``small'' $\epsilon_{\rm{in}}$ is characterized as the absence of interactions between two input errors---
nontrivial distributions of stockout states can be observed \textcolor{black}{(Fig. \ref{fig:dis})}, which have a minimum value of the stockout probability in the bulk. At the boundaries, we find local maximum values, and one of these values at stage-$s$ is essential for controlling the production rate.
From these observations, it can be concluded that in SP, for production systems with a large supply-error rate, it is more cost effective to increase the buffer size at the leaves than at downstream nodes in the assembly network.

Through simulations, we found that the effects of buffer size, $b$; degree of tree, $k$; and stage number, $s$, can be summarized as follows \textcolor{black}{(Figs. \ref{ssone} and \ref{fig:buwatto})}: (i) Increasing $k$ reduces particle flux, which can be absorbed by a sufficiently large $b$. However, compared with costs for maintaining large buffers, the effect is 
not significant; that is, the effect of flux fluctuations is not effectively addressed by solely increasing buffers (as long as  it is larger than the minimum size that allows steady flow ($b\geq 2$)).
(ii) a small stage number $s$ leads to the deterioration of flux in DP, 
while it contributes to increasing flux in SP.
(iii) In the presence of positive defect rate $\omega_{\rm{out}}$, large $b$ is effective in increasing DP flux,
while $k$ should be smaller when the system is not congested by particles.
These findings are worth considering as fundamental to improving understanding and designing the dynamics of real manufacturing systems.

When fluctuations are present, the stock level in each production block does not synchronize, which may reinforce production system instability. 
From a macroscopic view, among a group of factories (companies) that provide parts for end products, the bullwhip effect can be intensified by this uneven distribution of stocks, and vice versa. 
Note that the network-induced nonuniformity and bullwhip effects are substantially different phenomena; the former is characterized as horizontal inventory imbalance and the latter is characterized as a vertical inventory imbalance that appears only when the control of orders is based on the prediction of future demand. 

This paper is devoted to finding the fundamental properties of the process by setting a regular homogeneous network. 
However, it is also interesting to investigate the effects of network structure on production flow, heterogeneity of processing time, and buffer size, which will be the focus of future work. 
\textcolor{black}{Furthermore, it would be interesting to consider the technology diffusion in supply chain networks in the model \cite{Mc}}.
Although these changes influence the system dynamics, the presented interplay among various types of fluctuations, number of neighboring nodes, buffers are expected to be observed in general.

Although some problems remain, we believe that further study of the physical approach presented in this paper is promising and will facilitate effective understanding and management of production systems 
(e.g., understanding of how to place buffers effectively and characteristics of the response to environmental variations could be useful for addressing actual problems including the scheduling problem and cost optimization in factories). 
In this concept, the next goal is to show comparison with empirical data and application of statistical physics to the system.





\begin{acknowledgments}
We thank Hiroshi Takahashi, Kazuya Inaba, Kenta Yoshikawa, and Naokata Komuro for useful discussions. We would also like to acknowledge Ryosuke Nishi for his valuable comments.
This work was supported by JSPS Grants-in-Aid for Scientific Research (13J05086).
\end{acknowledgments}

\widetext
\appendix
\section{Production rate for $(s=1,b=1)$.}\label{asmall1}
When end products are manufactured by a single assembly ($s=1$) of $k$ parts and each buffer can contain one part at most at the same time ($b=1$), the production rate can be calculated thorough the average number of time steps required to accept $k$ different parts, $\bar{T}$.
Since parts provision to each buffer follows an identical and independent Poisson distribution, one obtains
\begin{eqnarray}
p(T=\tau) &=& P(T\leq \tau) - P(T\leq \tau-1)\nonumber\\
&=& (1-\epsilon^\tau)^k - (1-\epsilon^{\tau-1})^k,\\
\bar{T} &=& \sum_{\tau=0}^{\infty}{\tau\left[(1-\epsilon^\tau)^k - (1-\epsilon^{\tau-1})^k \right].}
\end{eqnarray}
Considering time for production, parts for an end product are output every $[\bar{T}+(1-\eo)^{-1}]$ time steps on average. Thus, the production rate is expressed as
\begin{equation}
q_{k} = \frac{1}{\bar{T}+\frac{1}{1-\epsilon_{\rm{out}}}}.
\end{equation}
For example, in Fig. \ref{ssone} (b), we use
\begin{equation}
\left\{
\begin{aligned}
q_1 & = \frac{(1-\ei)(1-\eo)}{2-\ei - \eo}\\
q_2 &= \frac{(1-\epsilon_{\rm{in}})(1+\ei)(1-\epsilon_{\rm{out}})}{2-\epsilon_{\rm{out}} + 2 \epsilon_{\rm{in}}(1-\epsilon_{\rm{out}})-\epsilon_{\rm{in}}^2}\\
q_4 & = \frac{1}{\frac{1+5\ei+3\ei^2+10\ei^3+2\ei^4+4\ei^5}{1+\ei+\ei^2-\ei^4-\ei^5-\ei^6} + \frac{1}{1-\eo}}
\end{aligned}
\right. .\label{qsm}
\end{equation}

\section{Probability distributions for $s=1$, $k=(1,2),$ and $b=(1,2)$ cases.}\label{asmall2}
When the number of states realized in a system is sufficiently small, we can 
calculate their probability distributions exactly.
Let $P_{\tau_1\cdots\tau_{k}}$ be the probability of finding a state $\{\tau_1,\cdots,\tau_k\}$,
where $\tau_i$ indicates the number of particles in the $i$th buffer.
With a probability vector  $\bm{P}=(P_{0\cdots 0},\cdots,P_{b\cdots b})^T$ and a transition matrix $T_{(b+1)^k\times (b+1)^k}$, the probability distribution is obtained 
through solving the following equation:
\begin{equation}
T_{(b+1)^k\times (b+1)^k} \bm{P} = \bm{P}.\label{ssseq}
\end{equation}
Using the obtained distribution, we can calculate particle flux as
\begin{equation}
q = (1 - \epsilon_{\rm{out}})\sum_{\prod_{1\leq i\leq k} \tau_i \neq 0} P_{\tau_{1}\cdots \tau_{k}}, \label{fluxccal}
\end{equation}
where the sum is taken over all possible configurations of particles, where at least one particle is stored in all the buffers.

In the following equations, explicit expressions of this equation for each $(k,b)$ are shown.
\subsection{$k=1,b=1$}
Eqs. (\ref{ssseq}) 
\begin{equation}
\begin{pmatrix}
\epsilon_{\rm{in}} & \beo\\
\bei & \epsilon_{\rm{out}}
\end{pmatrix}
\left(
    \begin{array}{c}
      P_0 \\
      P_1 
    \end{array}
  \right)
=\left(
    \begin{array}{c}
      P_0 \\
      P_1 
    \end{array}
  \right)
\end{equation}
and (\ref{fluxccal}) yield
\begin{equation}
q = (1-\epsilon_{\rm{out}})P_1=\frac{(1-\epsilon_{\rm{in}})(1-\epsilon_{\rm{out}})}{2 - \epsilon_{\rm{in}} - \epsilon_{\rm{out}}},
\end{equation}
which coincides with Eq. (\ref{qsm}).
Note that we use abbreviations $\bei = 1 - \ei$ and $\beo =1 - \eo$ in the transition matrix. 

\subsection{$k=1,b=2$}
Eqs. (\ref{ssseq})
\begin{equation}
\begin{pmatrix}
\epsilon_{\rm{in}} & \epsilon_{\rm{in}}\beo & 0\\
1-\epsilon_{\rm{in}} &\bei \beo+ \epsilon_{\rm{in}}\epsilon_{\rm{out}} & \beo\\
0 & \bei\epsilon_{\rm{out}} &\epsilon_{\rm{out}}
\end{pmatrix}
\left(
    \begin{array}{c}
      P_0 \\
      P_1 \\
      P_2
    \end{array}
  \right)
=\left(
    \begin{array}{c}
      P_0 \\
      P_1 \\
      P_2
    \end{array}
  \right)
\end{equation}
and (\ref{fluxccal}) yield
\begin{equation}
q = (1-\epsilon_{\rm{out}}) (P_1+P_2)=\frac{
(1-\ei)(1-\eo)(1-\ei\eo)
}{
1-\ei\eo(3-\eo)+\ei^2\eo
}.
\end{equation}

\subsection{$k=2, b = 1$}
Eqs. (\ref{ssseq})
\begin{equation}
\begin{pmatrix}
\ei^2 & 0  & 0 & \beo\\
\bei \ei & \ei & 0 & 0\\
\bei \ei & 0  & \ei & 0\\
\bei^2 & \bei & \bei & \eo 
\end{pmatrix}
\left(\begin{array}{c}
      P_{00} \\
      P_{01} \\
      P_{10} \\
      P_{11}
    \end{array}\right)
=\left(\begin{array}{c}
      P_{00} \\
      P_{01} \\
      P_{10} \\
      P_{11}
    \end{array}\right)
\end{equation}
and  (\ref{fluxccal}) yield
\begin{equation}
q = (1 - \eo) P_{11} = \frac{(1+\ei)(1-\ei)(1-\eo)}{2  - \eo + 2 \ei(1-\eo)- \ei^2}.
\end{equation}

\subsection{$k=2,b=2$}
Eq. (\ref{ssseq}) is written as
\begin{equation}
\begin{pmatrix}
\ei^2& 0&0&0&\ei^2 \beo&0&0&0&0\\
\bei \ei&\ei^2&0&0&\ei\bei\beo& \ei\beo&0&0&0\\
0&\bei\ei&\ei&0&0&0&0&0&0\\
\bei\ei&0&0&\ei^2&\ei\bei\beo&0&0&\ei\beo&0\\
\bei^2&\ei\bei&0&\ei\bei&\bei^2\beo+\ei^2\eo&\bei\beo&0&\bei\beo&\beo\\
0&\bei^2&\bei&0&\ei\bei\eo&\ei\eo&0&0&0\\
0&0&0&\ei\bei&0&0&\ei&0&0\\
0&0&0&\bei^2&\ei\bei\eo&0&\bei&\ei\eo&0\\
0&0&0&0&\bei^2\eo&\bei\eo&0&\bei\eo&\eo
\end{pmatrix}
\left(\begin{array}{c}
      P_{00} \\
      P_{01} \\
      P_{02} \\
      P_{10} \\
      P_{11} \\
      P_{12} \\
      P_{20} \\
      P_{21} \\
      P_{22} 
\end{array}\right)
=\left(\begin{array}{c}
      P_{00} \\
      P_{01} \\
      P_{02} \\
      P_{10} \\
      P_{11} \\
      P_{12} \\
      P_{20} \\
      P_{21} \\
      P_{22} 
\end{array}\right).
\end{equation}
Though particle flux can be obtained from (\ref{fluxccal})
\begin{equation}
q= P_{11}+P_{12}+P_{21}+P_{22},
\end{equation}
its explicit expression is too complicated to be written here.

\section{Production rate for $b\rightarrow\infty$.} \label{infbuff}
We can also calculate the production rate when the buffer capacity is infinite. 
Differing from the previous section, we here focus on an arbitrarily long time $T'$ and  count the number of
parts provided to the boundary in that period, since the parts' inputs are never blocked.
When $T'$ is sufficiently large, the expected number of end products manufactured, $qT'$, coincides with 
the number of the part that is provided fewest times.
Thus, we evaluate the minimum number of parts provisions to bottom buffers.
We define $S_i$ to be the number of parts provided to buffer $i$ ($i=1,\cdots,k^s$) in a time interval $T'$,
and $S_{\rm{min}} = \min_{1\leq i\leq k^s}{\{S_i\}}$.
\begin{eqnarray}
P(S_{\rm{min}}\geq m)& =& \prod_{i=1}^{k^s}P(S_i\geq m) \\
&=& \left(1- \sum_{m=0}^{T}\binom{T}{m} \epsilon_{\rm{in}}^{T-m} (1-\epsilon_{\rm{in}})^m\right)^{k^s}
\end{eqnarray}
\begin{eqnarray}
q& =&\frac{1}{T'} \sum_{m=0}^{T'} m P(S_{\rm{min}} = m) \\
&=& \frac{1}{T'}\sum_{m=0}^{T'}m\left[P(S_{\rm{min}}\geq m) -P(S_{\rm{min}}\geq m+1)\right] \\
&=& \frac{1}{T'}\sum_{m=1}^{T'}\left(1- \sum_{l=0}^{m-1}\binom{T'}{l} \epsilon_{\rm{in}}^{T'-l} (1-\epsilon_{\rm{in}})^l \right)^{k^s}\\
&\simeq& \frac{1}{T'}\sum_{m=1}^{T'}\left(1- \int_{0}^{m-1}
\frac{
	\exp{\left(\frac{
					-(x-(1-\epsilon_{\rm{in}})T')^2
					}{
					2\epsilon_{\rm{in}}(1-\epsilon_{\rm{in}})T'}\right)
		}
	}{
	\sqrt{2\pi \epsilon_{\rm{in}}(1-\epsilon_{\rm{in}})T'}}dx \right)^{k^s}\label{ap1}\\
&\simeq&\frac{1}{T'}\sum_{m=1}^{T'} [U(m-(1-\epsilon_{\rm{in}})T')]^{k^s}\\
&\simeq&1-\epsilon_{\rm{in}}\label{ap2},
\end{eqnarray}
where $U(x)$ is a unit step function that returns $0$ if $x<0$ and $1$ otherwise.
Approximations (\ref{ap1})-(\ref{ap2}) are exact when $T'\rightarrow\infty$.
The obtained result $q=1-\epsilon_{\rm{in}}$ is consistent with the fact that 
the production flow is totally controlled by a single stockout path, to which a new material is 
provided with a rate $1-\epsilon_{\rm{in}}$.




\end{document}